# Unconventional order parameter induced by helical chiral molecules adsorbed on a metal proximity-coupled to a superconductor

*Tamar Shapira[1,&], Hen Alpern[&,1], Shira Yochelis[2], Ting-Kuo Lee,[3] Chao-Cheng Kaun,[4] Yossi Paltiel[2], Gad Koren[5], Oded Millo*,[1]*

[1] Racah Institute of Physics and the Center for Nanoscience and Nanotechnology, The Hebrew University of Jerusalem, Jerusalem 91904, Israel

[2] Applied Physics Department the Center for Nanoscience and Nanotechnology, The Hebrew University of Jerusalem, Jerusalem 91904, Israel

[3] Institute of Physics, Academia Sinica, Taipei 11529, Taiwan

[4] Research Center for Applied Sciences, Academia Sinica, Taipei 11529, Taiwan

[5] Department of Physics, Technion - Israel Institute of Technology, Haifa 32000, Israel

*Corresponding Author: milode@mail.huji.ac.il

[&]Tamar Shapira and Hen Alpern contributed equally to this work.

# ABSTRACT

Following our previous results, which provide evidence for the emergence of a chiral *p*-wave triplet-pairing component in superconducting Nb upon the adsorption of chiral molecules, we turned to investigate whether such an effect can take place in a proximal superconductor consisting of metal on superconductor bilayer. Note that in such proximity systems, correlated electron-hole (Andreev) pairs exist in the normal metal rather than genuine Cooper pairs. To that end, we used scanning tunneling spectroscopy (STS) on thin Au films grown *in-situ* on NbN (a conventional *s*-wave superconductor) before and after adsorbing *helical chiral,* alpha-helix polyalanine molecules. The tunneling spectra measured on the pristine Au surface showed conventional (*s*-wave like) proximity gaps. However, upon molecules adsorption the spectra significantly changed, all exhibiting a zero-bias conductance peak embedded inside a gap, indicating unconventional superconductivity. The peak reduced with magnetic field but did not split, consistent with equal-spin triplet-pairing *p*-wave symmetry. In contrast, adsorption of *non-helical chiral* cysteine molecules did not yield any apparent change in the order parameter, and the tunneling spectra exhibited only gaps free of in-gap structure.

# I. INTRODUCTION

Compelling experimental evidence for the emergence of triplet superconductivity at superconductor-ferromagnetic interfaces has been accumulated during the past 15 years (as reviewed in Ref. [1]), corroborating theoretical predictions [2–4]. In particular, various groups reported long-ranged spin-polarized supercurrents in Josephson junctions [5–8], over distances much larger than the expected coherence length in a ferromagnet, $\xi_F = \sqrt{\hbar D/2E_{ex}}$ [9], where $D$ is the diffusivity and $E_{ex}$ the exchange energy in the ferromagnetic material. Additional evidence was provided also by scanning tunneling microscopy and spectroscopy (STM-STS) of bilayers comprising superconducting and ferromagnetic materials, which exhibited long-range penetration of superconductor order into the ferromagnet, consistent with proximity-induced triplet superconductivity [10–12]. The tunneling spectra exhibited zero-bias conductance peaks (ZBCPs), signifying the appearance of unconventional order-parameter (OP) symmetry, as expected from theoretical considerations. Due to the required Fermionic anti-symmetrization, the "orbital" symmetry in the triplet state can be either even (*s-* or *d-*wave) or odd (*p-*wave), corresponding, respectively, to odd- or even-frequency superconductivity [4,13]. All these OPs can manifest themselves by the appearance of ZBCPs in the tunneling spectra, for certain tunneling directions, in particular the anisotropic sign-changing ones (*p-* and *d-*wave), for which the ZBCPs are more pronounced [14–16]. It should also be pointed out, and quite relevant to our present work, that strong evidence for triplet superconductivity penetrating the (pristine singlet) superconducting side of ferromagnetic-superconducting interfaces was also reported by us and by other groups [17–19].

In a recent work we have demonstrated another method for changing the surface order parameter, which does not rely on ferromagnetic materials [20]. We have shown, using STS, that the superconducting order parameter in Nb (a singlet, *s*-wave superconductor) is altered upon the adsorption of chiral helical molecules - alpha-helix polyalanine. After molecules adsorption, the tunneling spectra displayed ZBCPs embedded in gaps, and their overall shape conformed to a combination of *s*-wave and equal-spin triplet chiral *p*-wave pairing potentials. However, the possibility of chiral- *d*-wave induced OP (that can also be associated with a triplet-pairing state) could not be ruled out, a question that is addressed below. A second question that arises from this previous work is whether chiral molecules can induce triplet superconductivity in a conventional proximal superconductor, namely, in a metallic film coupled to an *s*-wave superconductor. We note that in a proximal superconductor there are no genuine Cooper pairs but rather correlated electron-hole pairs ("Andreev pairs") [21,22], and therefore it is not a trivial question. From the applicative viewpoint, the ability to induce triplet superconductivity in metallic interconnects is an important step towards the realization of superconducting spintronic devices. And finally, it is not yet clear whether molecular helicity (as in alpha-helix polyalanine) is essential for the OP modification, or that mere chirality suffices.

In the present work we demonstrate that chiral, helical molecules can modify also the OP symmetry of proximity-induced superconductivity. STS measurements were performed on Au films proximity coupled to NbN (a conventional singlet-pairing *s*-wave superconductor) before and after the adsorption of polyalanine alpha-helix chiral molecules (see Supplemental Material). Similar to the case of Nb, we find here also that upon molecules' adsorption ZBCPs embedded in gaps appeared in the tunneling spectra, rather than conventional proximity gaps that were observed before adsorption, suggesting the emergence of unconventional superconductor OP. The post-adsorption tunneling spectra could be well fitted to a superposition of *s*-wave and equal-spin triplet *p*-wave OPs. The ZBCP height was reduced with the application of perpendicular magnetic field, but this peak did not split, consistent with an equal-spin triplet chiral *p*-wave OP [23]. However, non-helical chiral cysteine molecules (see Supplemental Material) appear not to modify the OP symmetry in either NbN or Au/NbN bilayer films.

## II. METHODS

Bilayers of Au (10-15 nm thick) on NbN (70 nm thick) were prepared *in-situ* without breaking the vacuum using pulsed laser ablation deposition on (100) SrTiO$_3$ (STO) wafers of 10×10 mm$^2$ area. The NbN films were deposited from a Nb target under 30 mTorr flow of N$_2$ gas at 500 ºC, whereas the Au films were deposited at 120 ºC under vacuum. The third-harmonic of a Nd-YAG laser at 355 nm wavelength and 10 J/cm$^2$ fluence on the targets was used for the deposition of both type of films. X-ray diffraction measurements of the NbN film on (100) STO showed mostly the cubic phase with a-axis orientation and a = 0.433 nm. The samples were then inserted into a 1 mM solution of polyalanine alpha-helix molecules dissolved in ethanol for 24 h.

The STM-STS measurements were performed at 4.2 K using a home-built cryogenic STM equipped with a small superconducting solenoid capable of producing magnetic fields up to ~400 mT normal to the films and using non-magnetic Pt-Ir tips. The topographic images were measured with current and bias settings of 0.1 nA and ~50-100 meV, well above the proximity gap in Au, while the tunneling *dI/dV-V* spectra, which are proportional to the local DoS, were acquired with lower bias voltages of 5-10 mV, before disconnecting the feedback loop, in order to increase the sensitivity to fine spectral features at the gap region. Correspondingly, the tunnel junction resistance were in the range $5 - 10 \times 10^7 \Omega$, well within the tunneling regime.

## III. RESULTS AND DISCUSSION

Figures 1a and 1d portray the STM-STS measurement schemes before and after the alpha helix molecules adsorption, respectively. Prior to adsorption, the STM topographic image reveals steps of 2-6 Å, consistent with single or double atomic steps in Au(111) [24–26], separated by atomically flat regions (Fig. 1b), as evident from the topographic line height-profile presented in Fig. 1c. After adsorption, the surface morphology changes markedly, and the topographic image exhibits rough surface morphology, as depicted by Figs. 1e. The green curve in Fig. 1f presents a topographic height-profile taken along a line normal to a stack of nearly parallel-aligned tilted molecules (green lines in Figs. 1(d) and 1(e)). It exhibits surface corrugation 0.8-1 nm in amplitude and separation of ~2 nm between peaks, each one corresponding to a single molecule. This is consistent with the diameter of the polyalanine alpha-helix molecule (~1 nm) we used [27,28]. The blue curve presents a scan along one of these molecules, and its length, ~2.3 nm, suggests the molecule, which is ~3 nm long, is tilted with respect to the surface by ~40° (and tilts of up to 50° were found), typical for non-uniform self-assembled monolayers of similar molecules [29] It is worth mentioning that the molecules do not form a uniform self-assembled monolayer with a distinct orientation but are rather disordered with varying bonding angles, a fact that may cause the large shape variations of the measured spectra over the sample area. Importantly, at any given point the spectra were reproducible and all exhibiting ZBCPs suggesting unconventional superconducting OP.

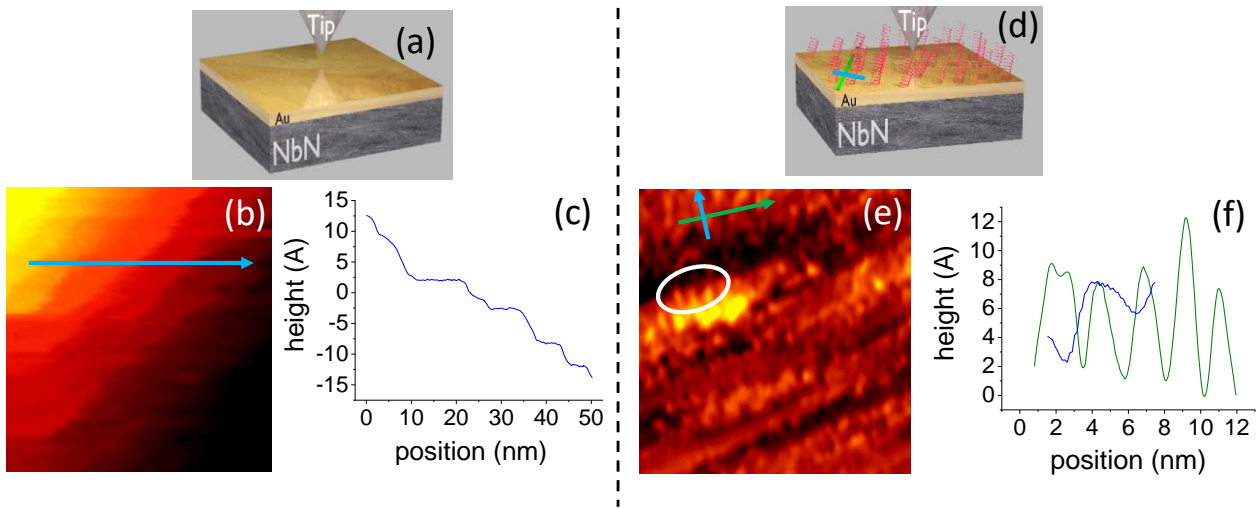

**Figure 1:** (a and d) Schemes of the STM-STS measurement on the bare and molecule-covered Au film deposited on NbN, respectively. (b and e) Typical topographic images measured on the bare (55×55 nm$^2$) and molecule-covered (32×32nm$^2$) Au film, respectively. (c and f) Line height profiles taken along the lines, and in the arrow directions, drawn in (b and e), respectively. The green curve in (f) was measured in a direction normal to a stack of parallel-aligned molecules, such as those encircled by the white ellipse, whereas the blue curve was measured

along a single molecule within the stack. The green and blue lines in (d) present schematically the cuts (with respect to the molecules) along which these green and blue height lines profiles were taken.

Figure 2 displays typical *dI/dV-V* tunneling spectra measured at 4.2 K before (a) and after (b-d) the helical chiral molecules adsorption. All of the tunneling spectra that were measured before adsorption are compatible with the DoS of a conventional *s*-wave proximal superconductor, as clearly visible in Fig. 2(a) that displays selected spectra measured on a 10 nm Au film deposited on NbN. All spectra exhibit a gap and coherence peaks that were well reproduced at any given location on the film, although their zero-bias conductance (depth) varied spatially between 0.34 to 0.5 of the normal (above gap) conductance, possibly reflecting spatial variations of the Au film thickness over the wafer and/or slightly different coupling of the Au to the NbN film. After adsorbing the polyalanine alpha-helix molecules, the *dI/dV-V* spectra markedly changed (Fig. 2b), displaying now ZBCPs embedded in gaps (usually lacking coherence peaks). Such spectra reflect the emergence of unconventional superconductivity, possibly with anisotropic sign-changing OP symmetry. Importantly, this new superconducting state remained also after exposing the sample to ambient conditions for three weeks, although the spectra features became much smaller (Fig. 2c). Similar behavior was observed for a 12 nm thick Au film deposited on NbN, as shown in figure 2(d): The proximity gaps of the bare Au/NbN film, such as those presented in the inset, transformed upon molecules adsorption into spectra portraying ZBCPs embedded in gaps or gap-like features, as presented in the main figure. Note that all spectral features, gaps and ZBCPs, measured on the bilayer with 12 nm thick Au cap layer are smaller than those measured on the one with 10 nm Au coating, and, importantly, all these superconducting-related features were not observed above Tc for all samples.

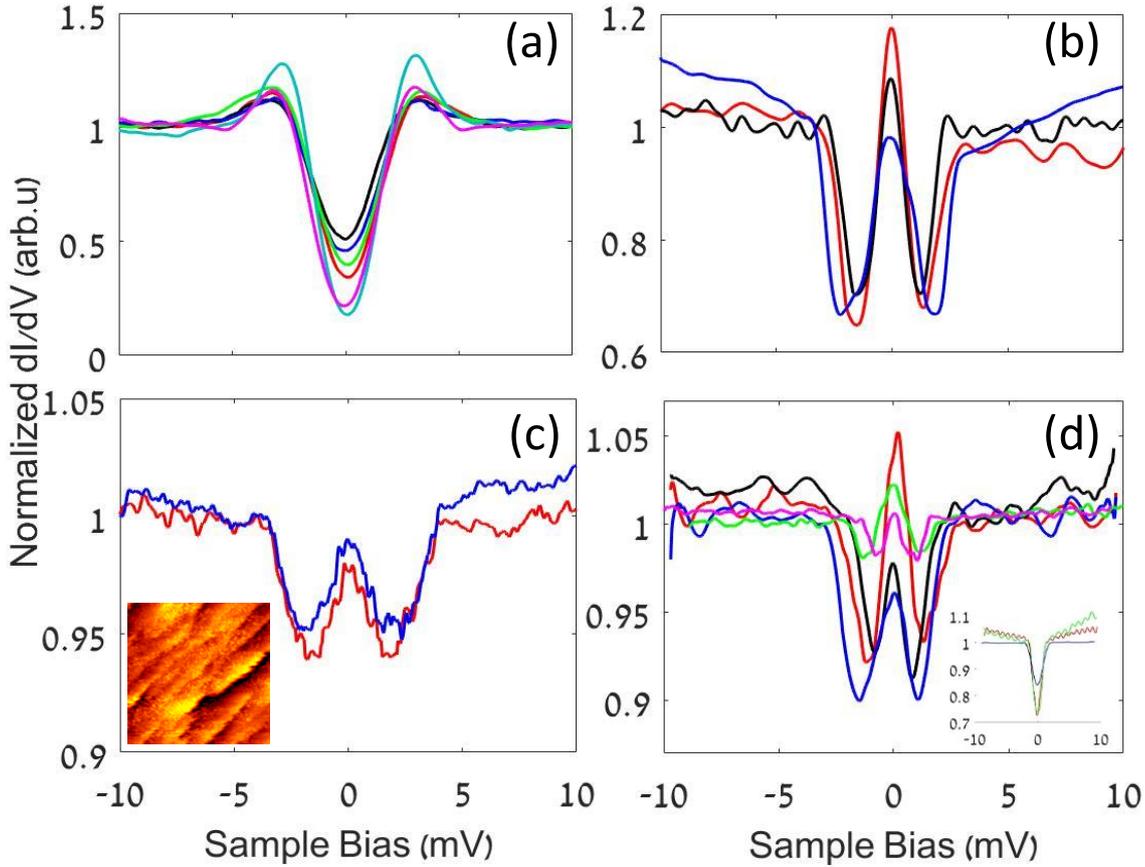

**Figure 2:** (a) Typical tunneling spectra measured on the bare surface of a 10 nm thick Au film deposited on 70 nm thick NbN layer, exhibiting proximity gaps. (b) Typical tunneling spectra measured after polyalanine alpha-helix molecules adsorption, showing the emergence of a ZBCP inside the gap. (c) Same as (b), but after 3 weeks in ambient conditions; the inset portrays a 100×100 nm$^2$ topographic image of the area where the spectra were acquired. (d) Same as (b), but for a 12 nm Au/NbN bilayer. The inset depicts tunneling spectra measured on the corresponding molecules-free Au/NbN bilayer.

In order to gain insight into the possible chiral-induced OP symmetries in the Au layer we fit the measured tunneling spectra to the Blonder, Tinkham and Klapwijk (BTK) model [30], modified to treat also other, non *s*-wave, OP symmetries, as detailed below and in the Supplemental Material. The standard parameters of such fits are the pairing amplitude (or energy gap), $\Delta_0$, the tunnel barrier strength, Z, and the Dynes lifetime broadening parameter [31], $\Gamma$. For unconventional OPs other fitting parameters may be required, as detailed below. All the theoretical spectra shown in Fig. 3 were calculated taking T = 4.2 K. We first address the spectra of the bare (pre-adsorption) bilayer (Fig. 2a), which appear to be compatible with conventional *s*-wave symmetry. Indeed, these spectra could be well fitted assuming a singlet-pairing *s*-wave OP, as shown in Fig. 3(a). The theoretical curve here was calculated with a gap of $\Delta_0 = 1.95$ meV, somewhat smaller than the value reported in Ref. [32] for Au/NbN bilayers, $\Gamma = 0.12$ meV and barrier strength $Z = 15$, well within the tunneling regime ($Z > 1$) [30]. The gaps extracted from fits to all spectra measured on the bare Au surface varied between 1.9 to 2 meV. Two more spectra showing these boundary gap values are presented in the Supplemental Material, Fig. S1

As discussed above, upon the adsorption of the chiral-helical molecules unconventional superconductivity emerges in the Au-NbN bilayer, as manifested by the appearance of ZBCPs inside gaps. Since the BTK model for *s*-wave superconductors cannot reproduce such results (even for very small Z values) [30], we tried fitting spectra such as those presented in Figs. 2b and 2c to extended BTK formalism, which include components associated with sign-changing OP symmetries, in particular *p*-wave [14,16] and *d*-wave [15], both giving rise to zero-energy Andreev bound states. We note in passing that the appearance (and magnitude) of ZBCPs in the tunneling spectra measured on *d*-wave or *p*-wave superconductors depend on the tunneling direction with respect to the crystallographic (and the corresponding pairing-potential) orientations. However, in the case of proximity-induced unconventional superconductivity, the crystallographic orientation role appears to be negligible, as ZBCPs appeared also in systems where the pristine superconductor was isotropic in nature [18,19].

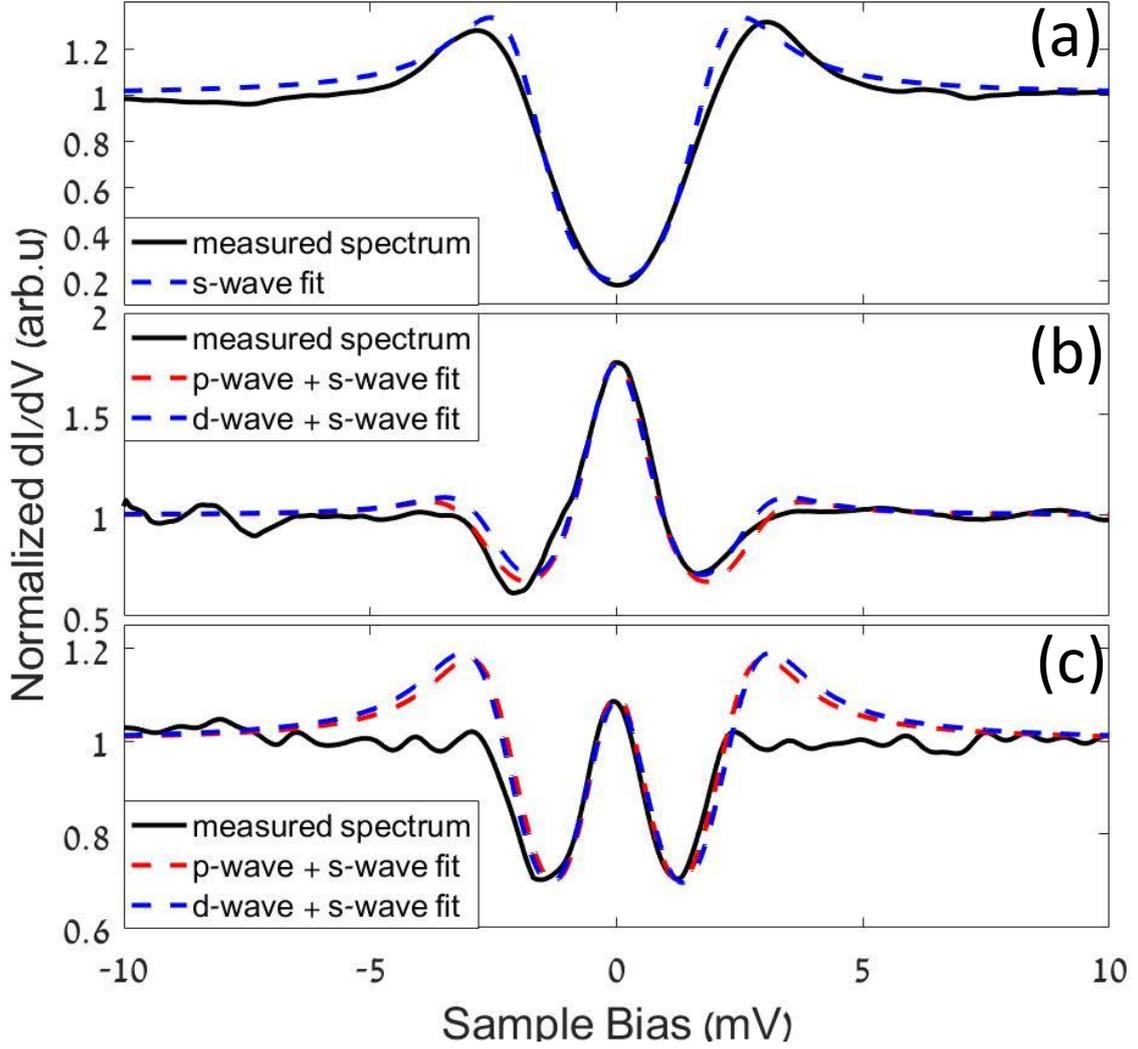

**Figure 3:** Theoretical fits to selected experimental dI/dV-V tunneling spectra (black curves). The theoretical spectra were calculated using the extended BTK formalism as detailed in the text and Supplemental Material. (a) Before polyalanine alpha-helix molecules adsorption; the theoretical spectrum (blue dashed curve) was calculated using only an *s*-wave component. (b and c) After adsorption; here the theoretical spectra were calculated assuming a combination of *s*-wave and *d*-wave (blue dashed line) and *s*-wave and chiral *p*-wave (red dashed line), as detailed in the text.

Two selected post-adsorption spectra manifesting non-conventional superconductivity together with corresponding fits are displayed in Figs. 3(b) and 3(c). The dashed red curves were calculated using a combination of *s*-wave and chiral *p*-wave ($p_x + ip_y$) OP symmetries, where the relative weight of the *p*-wave component is 70% in 3b and 46% in 3c. The specific chiral *p*-wave pairing-potential we used in the fits, which yielded best results, corresponds to an equal-spin (m=1) state which breaks time-reversal symmetry [14]: $\Delta_{\uparrow\uparrow}(\theta,\phi) = \Delta_0 \sin(\theta)e^{i\phi}$. Here, $\theta$ and $\phi$ are the polar and azimuthal angles, respectively. We have also tried fitting our data to chiral m=0 *p*-wave or non-chiral $p_x$-wave OPs but could not obtain reasonable fits. The blue dashed curves represent a weighted sum of *s*-wave and $d_{x2-y2}$-wave simulations. Note that a *d*-wave OP can still be associated with a triplet state, provided odd-frequency superconductivity occurs. Since we are not aware of any *d*-wave pairing-potential associated with triplet superconductivity, we used here the "standard" $d_{x2-y2}$-wave one, $\Delta = \Delta_0\cos(2\alpha)$; α being angle between the tunneling direction and the pairing-potential lobe. The relative contributions of the *d*-wave OP in the fits are 65% (42%) in 3b (3c), somewhat smaller than the corresponding *p*-wave weights. In spite of the good fits shown in Fig. 3b and to the ZBCP and dips (but not to the coherence peaks) in Fig. 3c, all with relatively small Dynes parameter, Γ ~ 0.08, there is a puzzle regarding the large energy gaps used in the fits, ~2.5 meV, which is comparable to the highest (tunneling) gaps reported for pure NbN films [33]. We note here that in many of the unconventional spectra we measured (which showed ZBCPs) the coherence peaks were strongly suppressed. This may possibly be due surface disorder which affect is known to affect these spectral features more than the gaps [34]. Fits to other spectra yielded gap values in the range 2.3-2.8 meV; two more spectra showing gaps of 2.35 and 2.8 meV are presented in Fig. S2. Possibly, the quasi-particle distribution on the Au surface was locally modified by charge exchange with the chemisorbed molecules in a way that increases the gap, as demonstrated by Blamire et al. for Nb-Al-Nb tunnel junctions [35]. This effect can be further enhanced in Au as a result of the small attractive pair interaction that was found to exist in Au [36]. A similar effect was observed in the case of the same chiral molecules adsorbed on Nb [20]. It should be noted here, however, that the opposite scenario, of gap reduction due to quasiparticle distribution modification, is more common [31,37–39], (see below). Spectra showing ZBCP can also manifest odd-frequency s-wave triplet superconductivity [2,40–42] and thus we cannot rule out this scenario, but we note that the STS data measured on Au/Nb/Ho presented in ref. [18] showed much smaller peaks associated with this OP.

As evident from Figs. 3a and 3b, the fits to the combinations of *s+d* and *s+p* OPs are equally good, and both were performed with similar parameters. Thus, we cannot use them in order to decide whether the additional chiral-induced OP is even-frequency chiral *p*-wave or odd-frequency *d*-wave. We therefore turned to investigate the effect of magnetic field on the tunneling spectra. The ZBCP in a *d*-wave superconductor was shown to split upon the application of magnetic fields even as low as 200 mT [43], whereas that related to an equal-spin triplet chiral $p_x + ip_y$ superconductor is expected to change only in magnitude [23,44]. The latter scenario was found in our magnetic field dependent results, presented in Fig. 4(a), suggesting that the unconventional superconducting component is indeed of equal-spin triplet chiral *p*-wave character. Note that the zero field ZBCP in figure 4(a) is wider than the ones embedded in gaps shown in Fig. 2 and the gap-feature is hardly observed; such a spectral shape, however, is possible for both *p* and *d*-wave OPs for different tunneling directions (with respect to the local pairing potential) [45,46]. It may also be due to different ratios between the conventional (*s*-wave) and unconventional components contributing to the spectrum. In that regard we should mention that this spectrum was acquired after the sample was exposed to magnetic field, which may have changed the molecular orientation and their coupling to the surface. A similar behavior was found in magneto-transport measurements on devices comprising chiral molecules [47]. However, this behavior under magnetic field *cannot* rule out the possibility that the ZBCP is associated with magnetic-like Yu-Shiba-Rusinov (YSR) states [48–50], *accidentally* centered at (or very close to) zero bias, because splitting should be too small to be detectable even at 400 mT ( ~0.05 meV), considering the observed peak width of 1.5 meV.

In order to further examine whether chirality (and possibly helicity, see below) is essential for the OP modification, and not merely the molecule-to-surface binding, we measured a reference (10 nm) Au /(70 nm) NbN bilayer after adsorbing non-chiral 1-decanethiol molecules. Fig. 4b shows that the OP symmetry is not modified due to this adsorption, namely, no ZBCPs appeared, although the gaps here, ~ 1.5 meV, are somewhat smaller than those measured on the bare Au/NbN bilayer.

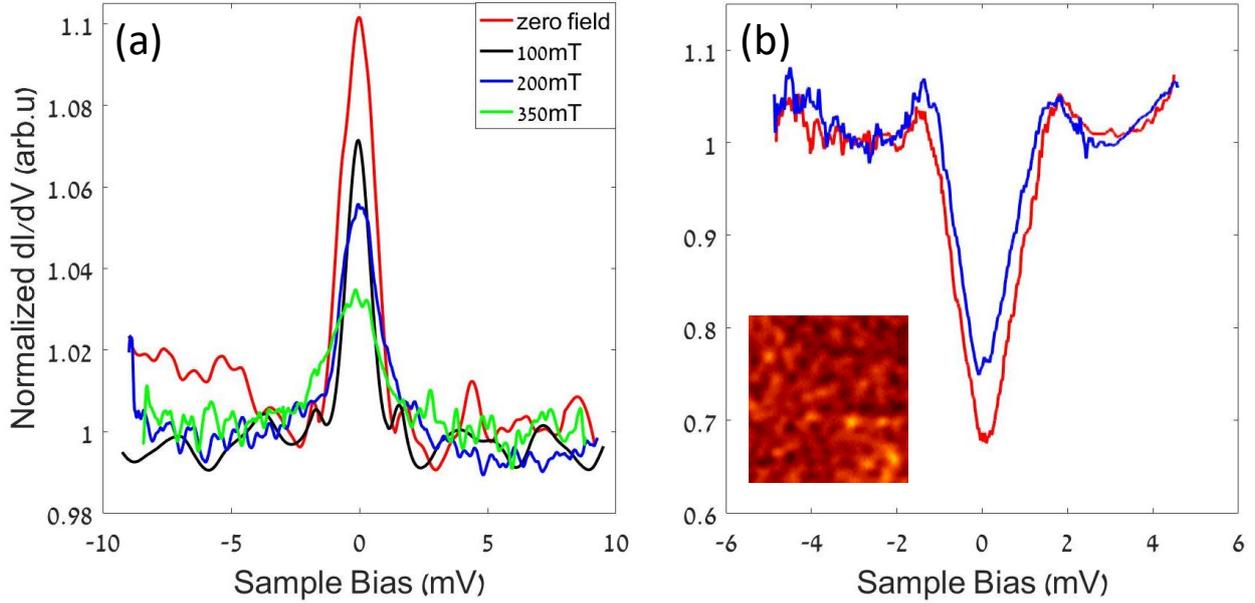

**Figure 4:** (a) Spectra measured on (10 nm)Au/(70 nm)NbN bilayer after *chiral* alpha-helix polyaniline molecules adsorption under different magnetic fields, as indicated. Note that the ZBCP does not split with magnetic field, making the *d*-wave scenario less probable. (b) Typical tunneling spectra measured on a (10 nm)Au/(70 nm)NbN bilayer after adsorption of the *non-chiral* 1-decanethiol molecules, for which ZBCP does not appear. A 30×30 nm² topographic image of the area where the spectra were measured, showing the molecular layer, is shown in the inset.

Next we turned to examine whether chirality by itself is sufficient for modifying the OP symmetry, or possibly molecular helicity is also required. To that end, we adsorbed chiral, but non-helical cysteine molecules on both bare NbN and on a 10nm Au/ 70nm NbN bilayer films by dipping them in distilled water solution for 10 min [55]. In both cases, the post-deposition spectra did not exhibit any signature for OP symmetry change; all showing gaps free of in-gap features (see Figs. 5 and S1). The gap values extracted by fitting the spectra presented in Fig. 5 were ~ 1.45-1.5 meV, comparable to those found for decanethiol-covered bilayers. Note that the gaps measured on some areas of the bare NbN (Fig. S1) are very shallow and exhibiting smaller gap values compared to those observed on the Au/NbN bilayer. This is probably due to an oxidized surface layer on the NbN (which can be about 1 nm thick) that does not develop during the in-situ Au/NbN bilayer growth, where the gold layer protects the NbN from oxidation. These results suggest that the helical geometry, and not only chirality by itself, may also be needed for the emergence of unconventional superconductivity. However, we cannot rule out at this stage that the absence OP change is due to the shorter length of the cysteine molecules (~0.7nm [56]) compared to the alpha-helix polyaniline molecules (3 nm). Further studies are needed in order to resolve this issue.

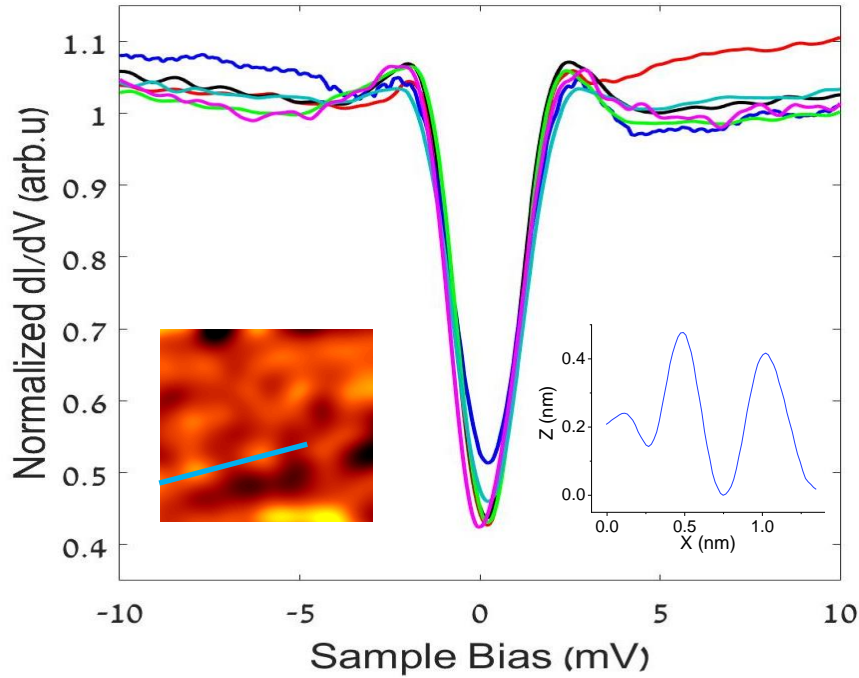

**Figure 5:** Tunneling spectra measured on (10 nm)Au/(70 nm)NbN bilayer after adsorbing non-helical chiral cysteine molecules. The spectra exhibit conventional gaps and no ZBCPs were found. The left inset presents a topographic image of the sample and the right one a cross-section taken along the blue line drawn in the image.

We still do not understand the mechanism underlying our findings, in particular if and how time-reversal symmetry is broken. As detailed in Ref. [20], one possible explanation relies on previous works demonstrating spin-selective transport through chiral molecules [57,58] and the consequent ability of these molecules to change the magnetization of Ni, either by passing current through them [59] or even just by adsorbing them on a Au-capped Ni surface [60,61]. It is thus plausible that a related spin-polarizing effect can act also on the Au/NbN surface, giving rise to a triplet superconductivity at the Au film. Note that strengths of the chiral potential and electrical dipole moment of the molecules are important parameters governing the spin-selectivity, which may not be strong enough in the non-helical short cysteine chiral molecules. Another possible scenario is related to theoretical predictions regarding the appearance of a triplet component in a two-dimensional superconductor [62], or at a superconductor/normal-metal interface [63], lacking inversion (or mirror) symmetry in the presence of strong Rashba type spin-orbit coupling (in the gold layer in the present study [64]). Under these conditions, mixing of excitations from the two spin bands takes place in a manner that makes spin a non-conserved quantity, yielding an intrinsic mixture of singlet and spin-aligned triplet-pair correlations. Our trilayer of chiral-molecule/Au/NbN resembles these model systems, in particular noting that the strong dipole moment of the adsorbed chiral molecules [65,66] can further enhances the strong Rashba spin orbit coupling at the Au surface.

## IV. CONCLUSIONS

In summary, we have demonstrated the appearance of unconventional superconductivity in the surface of thin Au films proximity-coupled to NbN due to adsorption of chiral-helical ployanaline alpha-helix molecules. This effect was clearly manifested by the emergence of zero-bias conductance peaks in the tunneling spectra measured on the Au cap layer upon molecules' adsorption, replacing the conventional BCS-like proximity gaps. The experimental tunneling spectra could be fitted equally well to two combinations of $s$-wave and either $d_{x2-y2}$-wave or equal-spin chiral $p$-wave triplet symmetries. However, the decrease, without splitting, of the ZBCP under the application of magnetic field points to the latter option, namely, that the chiral molecules induce an equal-spin triplet phase which breaks time-reversal symmetry in the proximal superconducting layer. Nevertheless, we cannot rule out the possibility that the chiral molecules induce odd-frequency triplet s-wave OP. These results may be useful for the development of practical chiral-based superconducting spintronic devices since, as we have shown here, the triplet superconductivity can be induced also in the metallic interconnect circuitry.

## ACKNOWLEDGMENTS

We thank Hadar Steinberg for helpful discussion. The research was supported in parts by the ISF F.I.R.S.T. program (grant # 687/16), the Niedersachsen Ministry of Science and Culture grant (O.M. and Y.P.), a grant from the Academia Sinica – Hebrew University Research Program (O.M., Y.P. and C.C.K.), and project number AS-iMATE-107-95 (T.K.L. and C.C.K.). O.M. thanks support from the Harry de Jur Chair in Applied Science.